\documentclass[preprint,11pt]{elsarticle}

\usepackage{graphicx}
\usepackage{amssymb}
 \usepackage{amsthm}
\usepackage{amsmath}

 \usepackage{lineno}

\journal{Nuclear Instruments and Methods in Physics Research A }

\begin{document}

\begin{frontmatter}


 %\title{Title\tnoteref{label1}}
 %\tnotetext[label1]{}
 %\author{\corref{cor1}\fnref{label2}}
 %\ead{zerguer@ipno.in2p3.fr}
 %\ead[url]{http://ipnweb.in2p3.fr}
 %\fntext[label2]{}
 %\cortext[cor1]{}
 %\address{Address\fnref{label3}}
 %\fntext[label3]{}

\title{Single electron response and energy resolution of a Micromegas detector}

 \author{T. Zerguerras \corref{cor1}\fnref{label1}}
 \ead{zerguer@ipno.in2p3.fr}
 %\ead[url]{http://ipnweb.in2p3.fr}
 %\fntext[label2]{}
 \cortext[cor1]{ Corresponding author Tel: +33 1 69 15 64 42 Fax: +33 1 69 15 50 01}
 \author[label1]{B. Genolini}
\author[label2]{V. Lepeltier$^{\dagger,}$}
\author[label1]{J. Peyr\'e}
  \author[label1]{J. Pouthas}
 \author[label1]{P. Rosier} 
 	
 \address[label1]{Institut de Physique Nucl\'eaire (UMR 8608), CNRS/IN2P3-Universit\'e Paris-Sud, F-91406 Orsay Cedex, France}
 \address[label2]{LAL, Universit\'e Paris-Sud, CNRS/IN2P3, Orsay, France}

%\author{}

%\address{}

\begin{abstract}
Micro-Pattern Gaseous Detectors (MPGDs) such as Micromegas or GEM are used in particle physics experiments for their capabilities in particle tracking at high rates. Their excellent position resolutions are well known but their energy characteristics have been less studied. The energy resolution is mainly affected by the ionisation processes and detector gain fluctuations. This paper presents a method to separetely measure those two contributions to the energy resolution of a Micromegas detector. The method relies on the injection of a controlled number of electrons. The Micromegas has a 1.6-mm drift zone and a 160-$\mu$m amplification gap. It is operated in Ne 95~\%-iC$\mathrm{_4}$H$\mathrm{_{10}}$ 5~\% at atmospheric pressure. The electrons are generated by non-linear photoelectric emission issued from the photons of a pulsed 337-nm wavelength laser coupled to a focusing system. The single electron response has been measured at different gains (3.7 10$\mathrm{^4}$, 5.0 10$\mathrm{^4}$ and 7.0 10$\mathrm{^4}$) and is fitted with a good agreement by a Polya distribution. From those fits, a relative gain variance of 0.31$\pm$0.02 is deduced. The setup has also been characterised at several voltages by fitting the energy resolution measured as a function of the number of primary electrons, ranging from 5 up to 210. A maximum value of the Fano factor (0.37) has been estimated for a 5.9~keV X-rays interacting in the Ne 95~\%-iC$\mathrm{_4}$H$\mathrm{_{10}}$ 5~\% gas mixture.
\end{abstract}

\begin{keyword}
Micromegas \sep Single-electron \sep Fano factor \sep Energy resolution
%%\PACS code \sep code
%%\MSC[2008] code \sep code
\end{keyword}

\end{frontmatter}

%\linenumbers

% SECTION 1: INTRODUCTION
%
%
\section{Introduction}
\label{sec:Intro}
Thanks to their very good resolution in position and capacity to work at high counting rates, Micro Pattern Gaseous Detectors (MPGD), like GEM \cite{sauli97} and Micromegas \cite{giomataris96}, have been studied to replace wire chambers for tracking in high energy physics experiments \cite{ketz02, koba07, kuden08}. More recently, those detectors have been used in nuclear physics experiments with heavy ions at energies ranging from several hundred keV up to a few MeV per nucleon. For example, a TPC (Time Projection Chamber) equipped with a four-GEMs structure was constructed for the study of the two-proton radioactivity \cite{blank08}. For future experiments with nuclear radioactive beams, R\&D activities have started on a TPC where the gas is both the target and the detection medium. This kind of detector must fulfil some specific conditions like the possibility of varying the pressure from a few tens of millibars to a few bars. An important feature is also the energy resolution that could be obtained on a large range of primary electrons depending on the ionisation power of the nuclei involved in the reactions. A small prototype with an active area of 16~cm$\mathrm{^2}$ has been constructed for dedicated studies on position and energy resolutions in the specific field of detection in nuclear physics. It is equipped with a Micromegas of a fairly large amplification gap (160~$\mu$m) in order to reach a sufficient gain at low pressures \cite{nakho08}. Before starting the studies at different pressures with different gas mixtures, we decided to investigate the energy resolution in more standard conditions, i.e. at atmospheric pressure and with a gas mixture that could allow a high dynamic range of gains without sparking. Different tests with Argon based gas mixtures showed that it is not possible to reach a sufficiently high gain with our large amplification gap and we chose Neon with a low amount of quencher: Ne 95~\% iC$\mathrm{_4}$H$\mathrm{_{10}}$5~\%.
The energy resolution depends on the fluctuations of the number of primary electron-ion pairs and the avalanche multiplication in the gas. The ionisation fluctuations can be quantified by the Fano factor \cite{fano47, alkha67} which depends on the gas and the energy of the primary particle (\cite{hurst78}-\cite{pansky97}). The gain fluctuations depend on the gas, the detector geometry and the applied electrical field \cite{byrne62, genz73}.  A powerful method to measure the gain fluctuations relies on knowledge of the detector Single Electron Response (SER). Then, by measuring the charge resolution for different numbers of primary electrons, the fluctuations of the primary charge can be estimated and a Fano factor deduced. The experimental set up is described in section \ref{sec:setup}. The method for the gain calibration independent of the SER measurements is presented in section \ref{sec:gaincalib}. It relies on the laser and the X-rays from a $\mathrm{^{55}}$Fe source. In section \ref{sec:single}, the SER at three different gains are shown together with their associated fits by Polya distributions. The factors affecting the energy resolution measured with the laser and the $\mathrm{^{55}}$Fe source are discussed in section \ref{sec:eneresol}.

%
%
% SECTION 2: EXPERIMENTAL SETUP
%
%

\section{Experimental setup}
\label{sec:setup}

\begin{figure}[htbp]
	\centering
		\includegraphics[width=12cm]{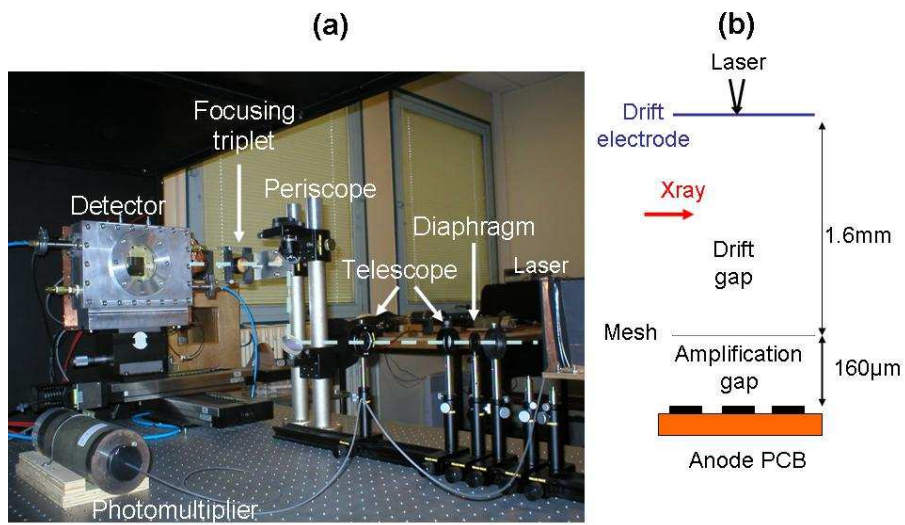}
	\caption{(a) View of the experimental setup (see text for details). the dotted line represents the laser beam axis (b) Schematics of the Micromegas detector.}
	\label{fig:figure1-setup}
\end{figure}

The experimental setup is similar to the one described in ref. \cite{zerguer07} and is shown in Fig. \ref{fig:figure1-setup}. The test-bench is composed of a pulsed laser, an optical line, and a 3D positioning system. The laser is a 337-nm wavelength pulsed Spectra-Physics VSL337 \cite{spectra08} and delivers pulses with a 4-ns FWHM duration, an energy of 120~$\mu$J, a pulse to pulse energy stability better than 4~\% in standard deviation, and a repetitive rate of 10~Hz. The laser beam is collimated by a diaphragm and a telescope, made of two spherical lenses and focused by a 60-mm focal length triplet. The spot size is smaller than 100~$\mu$m. In addition, a periscope is placed between the telescope and the triplet for mechanical reasons to adjust the beam height. The light intensity is tuned by means of commercial calibrated neutral density filters. The position and the focusing of the beam can be changed by moving the detector in the three directions of space with 1-$\mu$m accuracy controlled motors. 
The Micromegas detector has an active surface of 50$\times$50~mm$\mathrm{^2}$ and is enclosed in a vessel, equipped with a quartz entrance window. It is composed of a 1.6-mm drift gap and a 160-$\mu$m amplification gap, separated by a 400-lpi electroformed nickel mesh. The drift electrode is composed of a 0.5-nm thick Ni-Cr layer deposited on a quartz lamina. When the laser is focused on this electrode, the high energy density produces electrons by non-linear photoelectric emission processes \cite{anisi77}. The detector is operated with a Ne 95~\% iC$\mathrm{_4}$H$\mathrm{_{10}}$ 5~\% mixture at atmospheric pressure with gains varying from 10$\mathrm{^4}$ to 10$\mathrm{^5}$. The drift field is fixed to 1~kV/cm. A $\mathrm{^{55}}$Fe source is placed inside the vessel for absolute gain calibration. Different sets of pads and strips of various sizes are implemented on the anode PCB. In the present experiment, we use a set of nine 4$\times$4~mm$\mathrm{^2}$ pads.
The Micromegas mesh is connected to a voltage amplifier of gain 100 made with two NXP 5205A wide band amplifier integrated circuits \cite{nxpweb}. The anode pads readout is achieved by a GASSIPLEX chip \cite{santiard94}, which is a 16-channel multiplexed low noise preamplifier shaper. It works at a gain of 2.2~mV/fC, a peaking time of 1.2~$\mu$s and has a maximum dynamic range of 500~fC. The electronic noise is of 2~000 electrons RMS, which makes it possible to work at gains greater than 10$\mathrm{^4}$ for a low number of primary electrons. Digitisation (12-bit ADC), zero suppression and pedestal subtraction are performed using CAEN \cite{caenweb} modules (V550A and V551). The data acquisition system is triggered either by the mesh signal in runs with the $\mathrm{^{55}}$Fe source, or by a Photonis XP2282B photomultiplier \cite{photonisweb} connected to the laser through an optical fibre.

%
%
% SECTION 3: GAIN CALIBRATION
%
%
\section{Gain calibration}
\label{sec:gaincalib}

In principle the gain of the Micromegas detector can be deduced from the charge measurements in the single electron working mode. However, the standard calibration based on the 5.9~keV  X-rays of  $\mathrm{^{55}}$Fe  has been performed in order to crosscheck the results in the same gain domain. Unfortunately, the limited dynamic range of the GASSIPLEX circuit associated with the relatively high number of electrons given by the source does not make it possible to cover a wide gain range in the zone of interest. 
\begin{figure}[htbp]
	\centering
		\includegraphics[width=9cm]{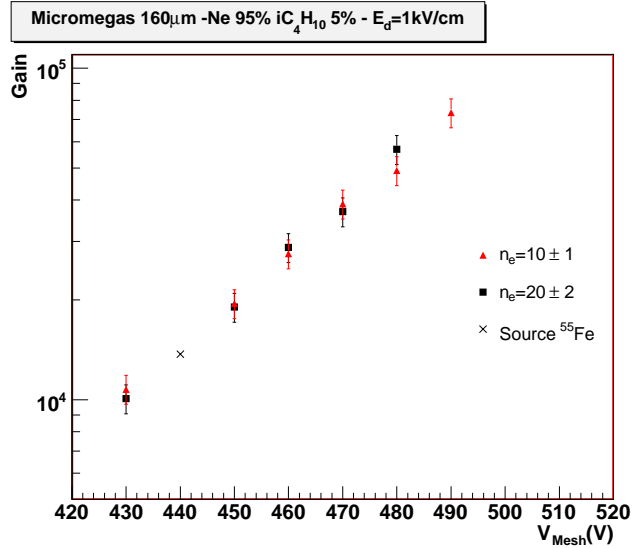}
	\caption{Gain of the Micromegas detector in Ne 95\% iC$_{4}$H$_{10}$ 5\%  as a function of the mesh voltage. Three sets of data are represented: $\mathrm {^{55}}$Fe source (cross), 10$\pm$1 triangles), and 20$\pm$2 (squares) electrons produced by focusing the laser on the drift electrode. }
	\label{fig:figure2-gain-calib}
\end{figure}
A method combining the source and the laser has been applied and the results are presented on Fig. \ref{fig:figure2-gain-calib}. The charge on the central 4$\times$4-mm$\mathrm{^2}$ pad is measured at a mesh voltage of 440~V with the $\mathrm{^{55}}$Fe source. It gives an absolute gain of 1.4 10$\mathrm{^4}$ with a number of primary electrons of 167. This number is the one obtained for a 5.9~keV X ray and for an ion-electron pair production mean energy of 35.3~eV in the Ne 95\% iC$\mathrm{_4}$H$\mathrm{_{10}}$ 5\% mixture ($W_{\mathrm{Ne}}$ = 36~eV and $W_{\mathrm{iC_4H_{10}}}$ = 23~eV being the mean energy for Neon and Isobutane \cite{leo94}). At the same voltage, the laser is focused on the drift electrode in front of the central pad. From the gain value obtained with the source, the number of primary electrons can be deduced. Then it can be adjusted at a much lower level than the one provided by the source. Measurements were performed with two values of primary electrons (10$\pm$1 and 20$\pm$2) and gave compatible results. The mesh voltage has been varied from 430~V to 490~V, i.e gains from 10$\mathrm{^4}$ to 7.10$\mathrm{^4}$.

%
%
% SECTION 4: SINGLE-ELECTRON MODE
%
%
\section{Single-electron mode}
\label{sec:single}

The method applied to reach the single electron mode is similar to the one used for photomultipliers \cite{photonis02}. The pulsed laser light intensity is attenuated to a regime in which the probability of producing one electron is much greater than the probability of giving more than one. In the present setup, the laser light intensity is attenuated by a factor of 2~000. In those conditions, only 7~\% of the events are considered as non-zero events (charge greater than 10$\mathrm{^4}$ electrons, which is 5 times the RMS noise) and the probability of obtaining more than one electron can be considered as low (0.5~\%). The charge spectra measured on the Micromegas central pad at five different biases (470 to 510~V) are shown in Fig.3. Experimental and theoretical studies in proportional gas counters \cite{genz73} showed that the gain fluctuations can be described by two models depending on the electric field to pressure ratio $R$, the gas characteristics and the detector geometries. For low $R$, the probability of ionisation by an electron collision within an avalanche can be assumed to be independent of its past history: the SER is exponential according to the Furry law. For larger $R$ values, measured SER showed pronounced maxima. This departure from the Furry distribution means that the ionising collision probability cannot be supposed independent of the path travelled from the previous ionisation: the SER follows a Polya distribution. Concerning MPGDs, the SER has been measured and fitted by Furry laws at moderate \cite{bondar07} and high \cite{buzulu00,vavra02} gains with stacks of GEMs. It has been fitted by a Polya for a triple GEM working at a gain of 4.1 10$\mathrm{^5}$ \cite{buzulu00}. The SER has also been measured with a Micromegas at gains greater than 10$\mathrm{^6}$ and follows a Polya distribution \cite{derre00}. Each measured SER clearly shows a maximum and can be fitted by a Polya distribution:

\begin{figure}[htbp]
\includegraphics[width=6.2cm]{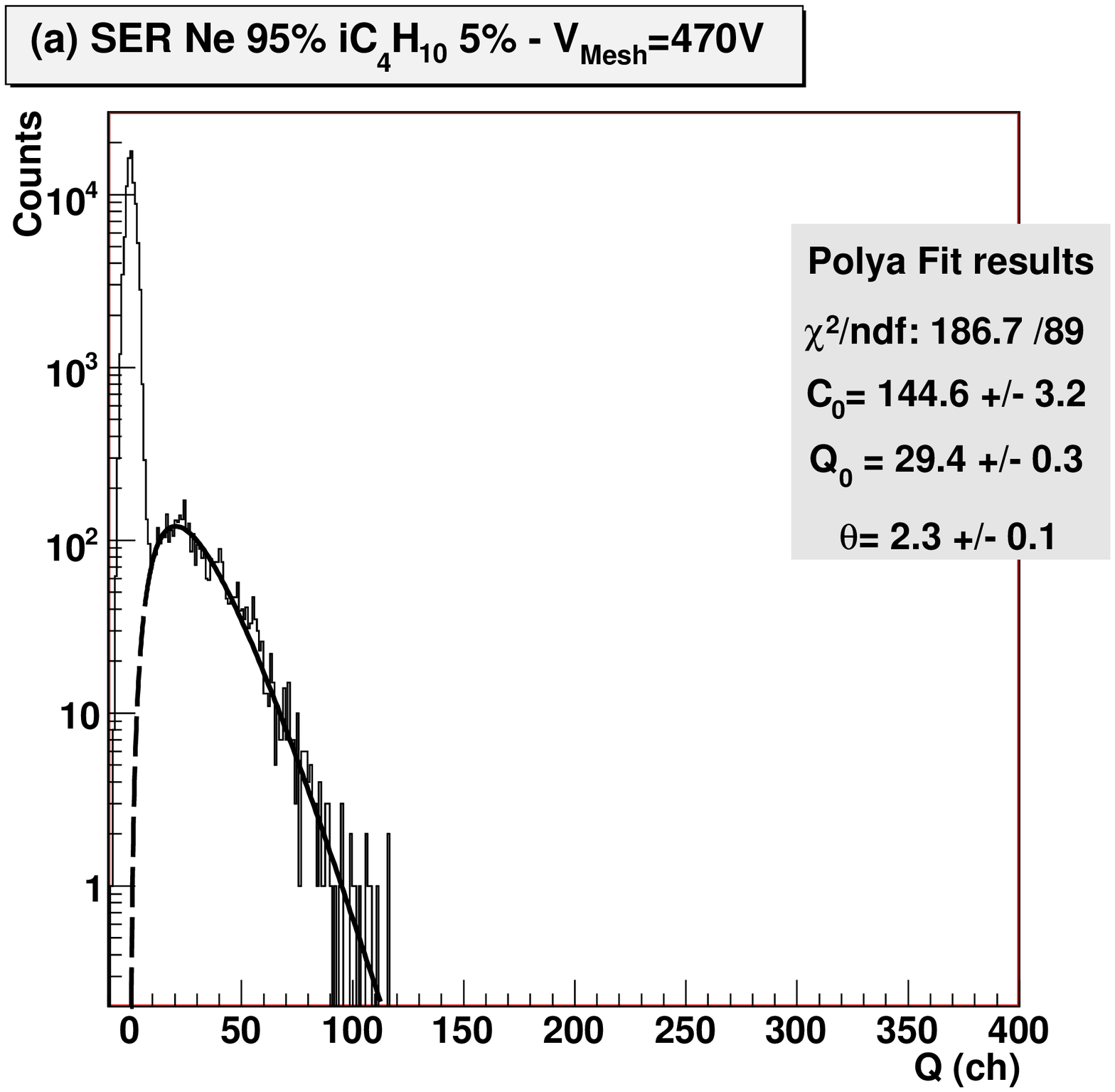}
\includegraphics[width=6.2cm]{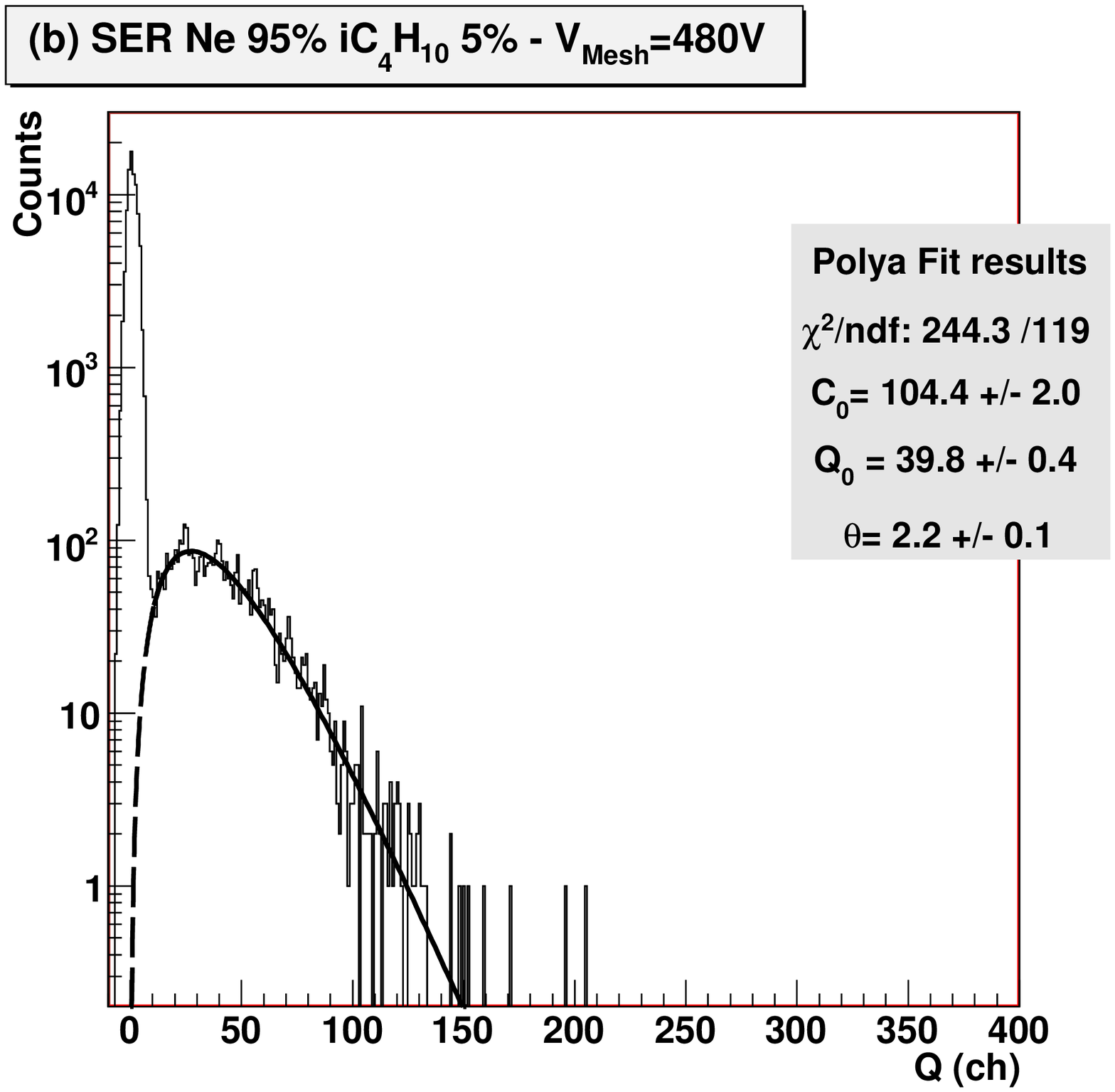}
\includegraphics[width=6.2cm]{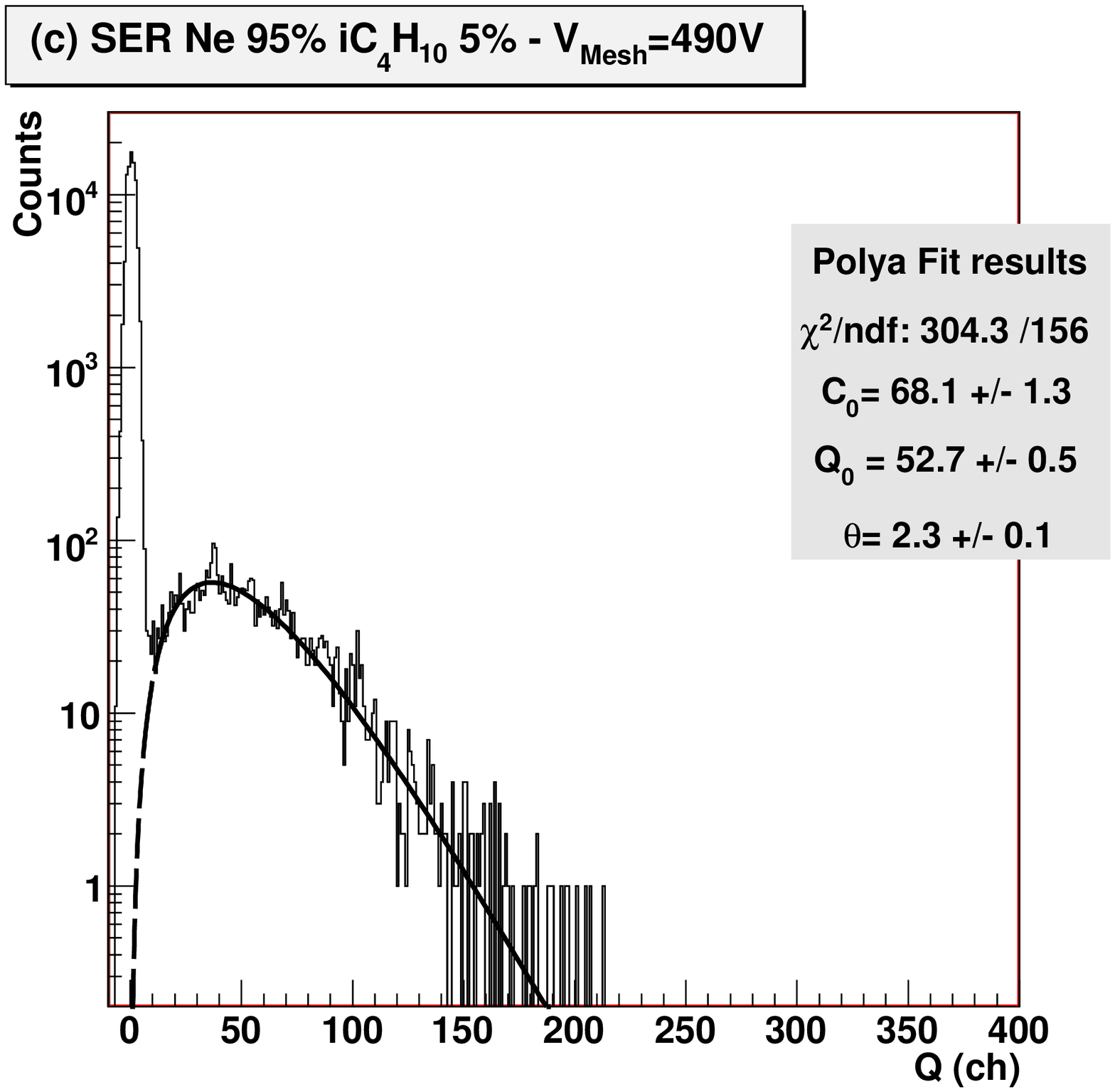}
\includegraphics[width=6.2cm]{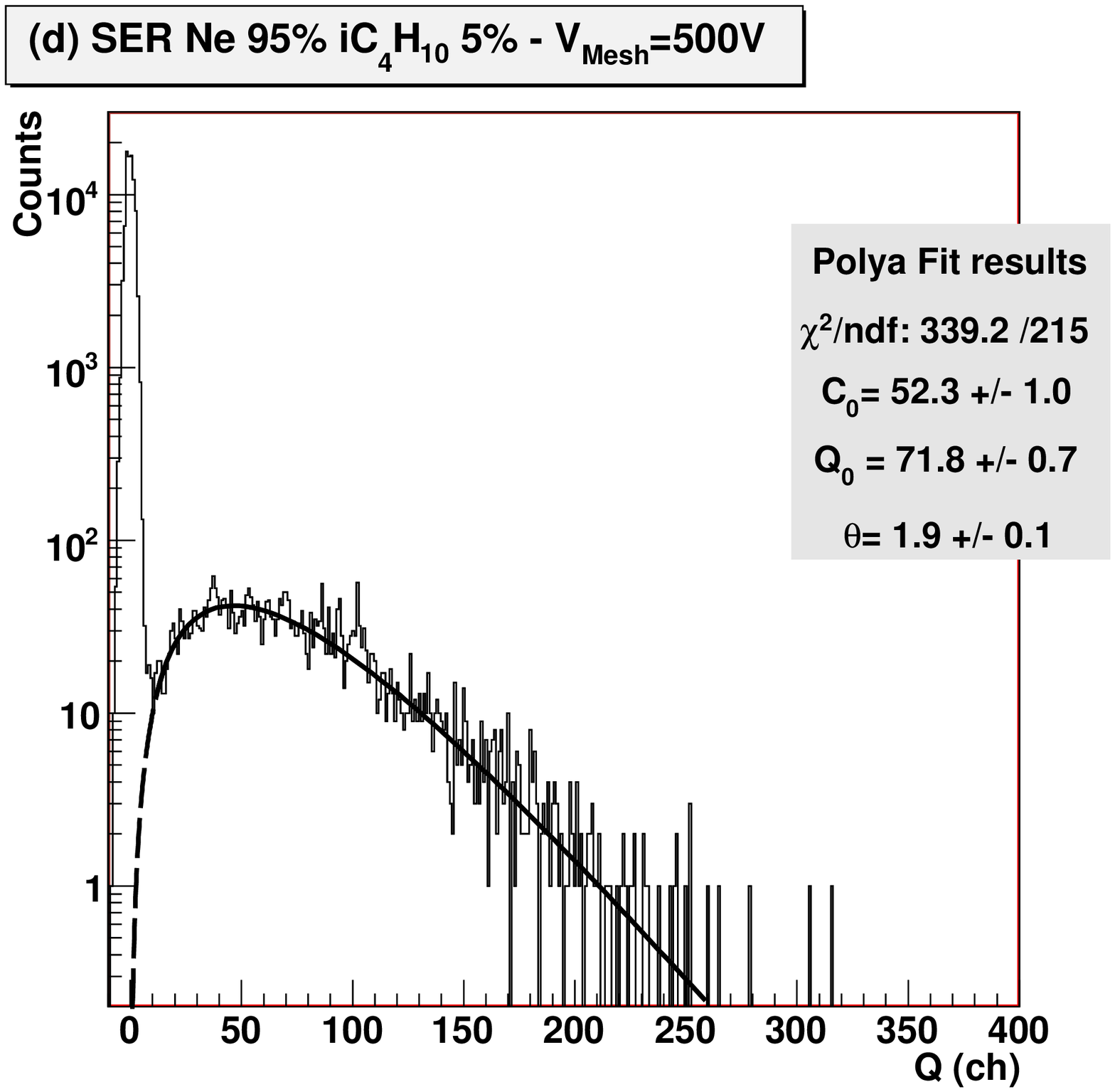}
\includegraphics[width=6.2cm]{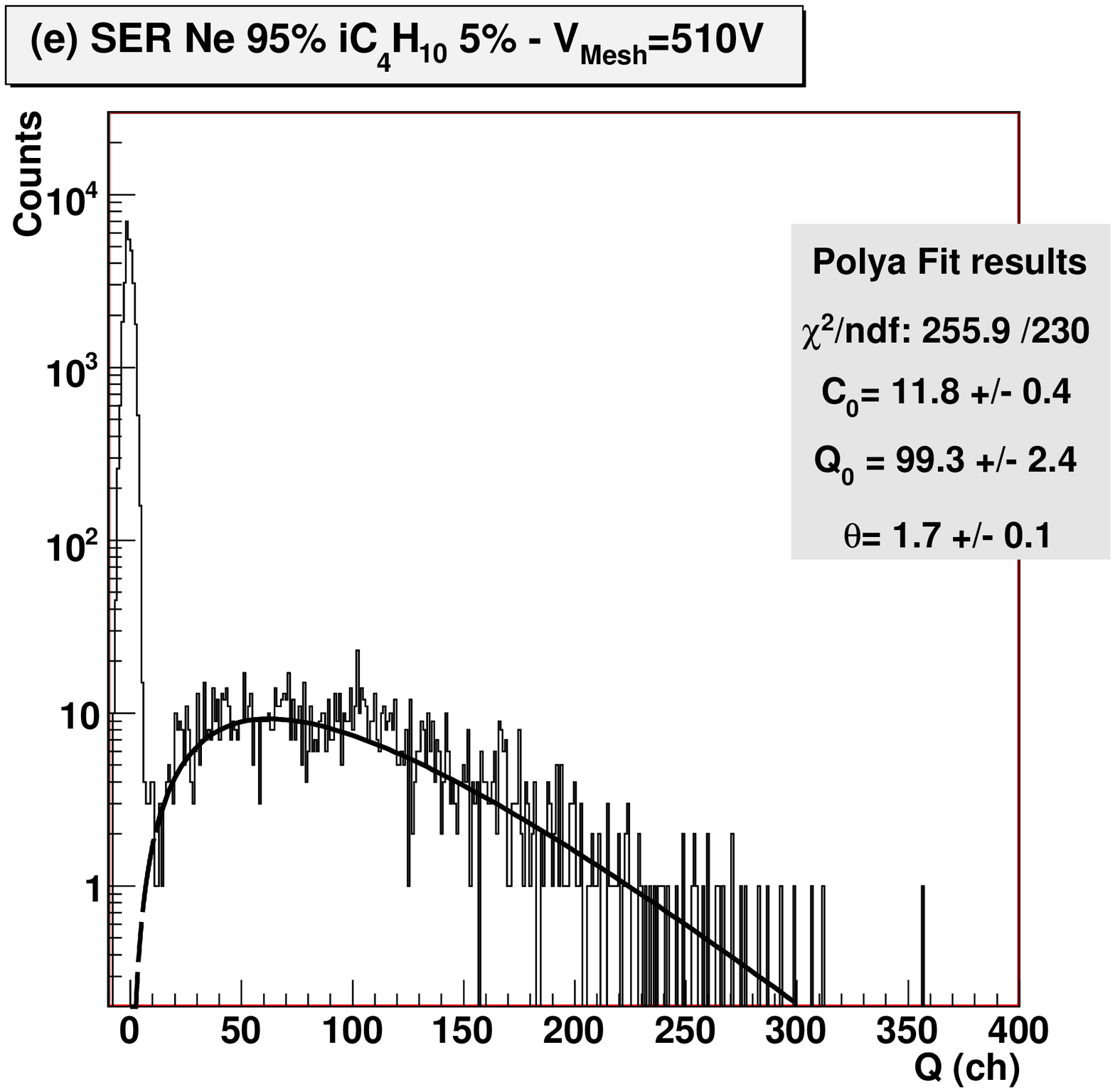}
\caption{Single-electron response (SER) of the Micromegas detector measured on one pad in Ne 95\% iC$_{4}$H$_{10}$ 5\% . Each spectrum is fitted with a Polya function (plain line). The dotted line is the extrapolation to zero of the fitted function. }
\label{fig:figure3-ser}
\end{figure}

\begin{eqnarray} \label{eqn:polya}
\mathrm{P}(Q) & = & C_{0} \frac{(1+\theta)^{1+\theta}}{\Gamma(1+\theta)} \left(\frac{Q}{Q_0} \right)^{\theta} \exp \left[-(1+\theta)\frac{Q}{Q_0} \right] 
\end{eqnarray}
where $C_{0}$ is a constant, $Q_{0}$ the mean charge and $\theta$ the parameter of the Polya.
%
%
% TABLE 1
%
%
\begin{table}[htbp]
\centering
\begin{tabular}{|c|c|c|}
\hline
Mesh voltage (V) & Gain (calibration) & Gain (single electron) \\
\hline
470 & 3.7 10$^{4} \pm0.3$ & 3.6 10$^{4}$ \\
\hline
480 & 5.0 10$^{4} \pm0.5$ & 4.6 10$^{4}$ \\
\hline
490 & 7.0 10$^{4} \pm0.7$ & 6.0 10$^{4}$ \\
\hline
\end{tabular}
\caption{Comparisons of the gains measured by the calibration (see section 3) and in single-electron mode.}
\label{table:tab1}
\end{table}
From the mean charge $Q_{0}$ , a gain value is deduced for each mesh voltage and for the three lower ones (470, 480 and 490~V) they are compared to the calibration values obtained in section 3 (see table \ref{table:tab1}). There is a good agreement between the two calibrations based on the 167 electrons of the $\mathrm{^{55}}$Fe source or deduced from the SER. Moreover, at those gains (3.7 10$\mathrm{^4}$, 5.0 10$\mathrm{^4}$ and 7.0 10$\mathrm{^4}$), the $\theta$ parameter does not depend on the gain: $\theta =2.2\pm0.2$. Even if no direct comparison can be made (different detector and different gas), it is encouraging to have a result similar ($\mathrm{\theta=2}$) to the one obtained with a triple-GEM operated in Ar+CH$\mathrm{_4}$ 1.3~\% at a gain of 4.1 10$\mathrm{^5}$ \cite{buzulu00}. The measurements made at the two larger bias voltages (500 and 510~V) give gains which are respectively less by 10~\% and 15~\% outside the error bars than the ones expected from the fit and extrapolation of the calibration results of section \ref{sec:gaincalib}. For those two gains, the parameter $\theta$ has smaller values: $\mathrm{\theta= 1.9}$ and 1.7 for 500~V and 510~V mesh bias respectively. The discrepancies could be explained by non linearity in the gas multiplication process particularly when a low proportion of quencher is used \cite{bronic96}. Anyway, it has been decided to only retain in further studies only the three SER results for which the agreement with the gain calibrations is satisfactory.

In detectors, an important feature is the gain fluctuations which can be expressed by the relative gain variance $f$ of the SER. If the SER is fitted by a Polya distribution, the relative gain variance is simply deduced from the parameter $\theta$ by \cite{knoll89}:

\begin{equation} \label{eqn:variance}
f= \frac{1}{1+\theta}
\end{equation}
From our measurements (biases at 470, 480 and 490~V) and the associated Polya fit, we have the same value for $\mathrm{\theta = 2.2\pm0.02}$ and thus the same for $f=0.31\pm0.02$. The gain variance measured in our operative conditions is two times lower than the one usually obtained in MWPC ($f$=0.7) which is sometimes taken as the parameter for the gain fluctuations in some MPGD's models \cite{koba06}. In \cite{derre00}, the SER has been measured in a Micromegas of 100~$\mu$m amplification gap with He-Isobutane 4\% and 10\% gas mixtures for gains above 10$\mathrm{^6}$. The spectra fit well with a Polya distribution, but the gain variance is double of that of our present result. However our value of $f$ is in good agreement with the one ($f$=0.33) obtained with a triple GEM operated in Ar+CH$\mathrm{_4}$ 1.3\% at a gain of 4.1 10$\mathrm{^5}$ \cite{buzulu00}.
Let us consider our two results rejected for calibration considerations. The detector was working at higher gains (biased at 500 and 510~V) and the Polya fit gave lower values for $\theta$ (1.9 and 1.7) and thus larger values for $f$ (0.34 and 0.37 respectively). It clearly indicates that avalanche fluctuations depend on the gain and increase at high gain values. That was already observed in parallel-plate proportional counters \cite{agraw88}. If the SER has to be used for energy resolution purposes, it must be measured at gains close to the one at which the detector is operated in the experiment.

%
%
%SECTION 5 : DETERMINATION OF THE ENERGY RESOLUTION
%
%
\section{Determination of the energy resolution}
\label{sec:eneresol}

The energy resolution is given by the charge resolution. The relative charge resolution $\sqrt{\mathrm{var}(Q)}/Q$ is related to the number of primary electrons $n_{e}$ by:

\begin{equation} \label{eqn:varqeq2}
\frac{\mathrm{var}(Q)}{(\mathrm{E}(Q))^2}= \frac{f}{\mathrm{E}(n_e)}+ \frac{\mathrm{var}(n_e)}{\mathrm{E}^2(n_e)} + \left(\frac{\sigma_I}{G}\right)^2 \frac{1}{\mathrm{E}^2(n_e)}
\end{equation}
where $f$ is the relative gain variance, $G$ the detector gain and  $\sigma_I$ the standard deviation of the noise at the electronics input. This formula and its conditions of application are discussed in Appendix \ref{app:appa}. In the current configuration, the electronic noise is 2~000 electrons RMS whereas the gain is always greater than 10$\mathrm{^4}$. Since $f =0.31$ and $\mathrm{(\sigma_I/G)^2 \leq 0.04}$, the term $(\sigma_I/G \mathrm{E}(n_e))^2$ is negligible compared to $f/\mathrm{E}(n_e)$, even for low number of electrons.

When the electrons are generated only by ionisation in the gas, as is the case with the X-rays produced by a $\mathrm{^{55}}$Fe source, the variance of $n_e$ can be expressed by the following formula:

\begin{equation} \label{eqn:fanosource}
\frac{\mathrm{var}(n_e)}{\mathrm{E}^2(n_e)}=\frac{F}{\mathrm{E}(n_e)}	
\end{equation}
where $F$ is a constant called the Fano factor \cite{fano47, alkha67}. It depends on the interactions of the particle with the gas under consideration.

When the primary electrons are generated by laser photons interacting with a metal, the variations of $n_e$ depend on the laser energy fluctuations and on the statistics of photon absorption:

\begin{equation} \label{eqn:fanolaser}
\frac{\mathrm{var}(n_e)}{\mathrm{E}^2(n_e)}= a^2 + \frac{F_{las}}{\mathrm{E}(n_e)}	
\end{equation}
where $a$ is a parameter related to the energy resolution of the laser, and $F_{las}$ is equivalent to a Fano factor associated with the electron emission from the photons. The details of the calculation and the conditions of application are also given in Appendix \ref{app:appa}.

These expressions of the variance of $n_e$ have been used to characterise our Micromegas detector operating with Ne 95~\% iC$\mathrm{_4}$H$\mathrm{_{10}}$ 5~\% gas mixture.

%\newpage
%
%
% SUBSECTION 5.1: MEASUREMENT  WITH THE LASER
%
%
\subsection{Measurement with the laser}
\label{subsec:fanolaser}

The expression of the charge resolution measured with the laser is obtained by applying (\ref{eqn:fanolaser}) to (\ref{eqn:varqeq2}):

\begin{equation} \label{eqn:chargeresollas}
\left(\frac{\sigma_Q}{Q} \right)= \frac{\sqrt{\mathrm{var}(Q)}}{\mathrm{E}(Q)}=\sqrt{a^2+\frac{f+F_{las}}{\mathrm{E}(n_e)}}
\end{equation}

\begin{figure}[htbp]
	\centering
		\includegraphics[width=9cm]{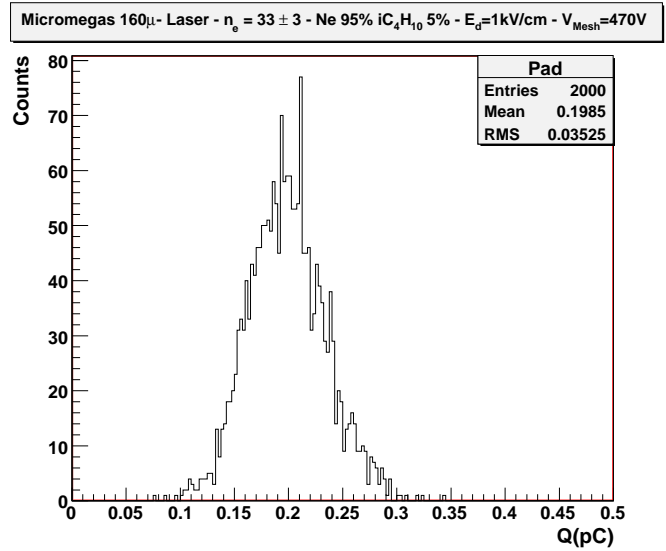}
	\caption{Charge induced on one pad of the Micromegas detector by 33$\pm$3 electrons produced by focusing the laser on the drift electrode. The gas mixture is Ne 95\% iC$_{4}$H$_{10}$ 5\%, and the mesh voltage 470~V. }
	\label{fig:figure4-qtot-ne33-vg470}
\end{figure}
\begin{figure}[htbp]
	\centering
		\includegraphics[width=9cm]{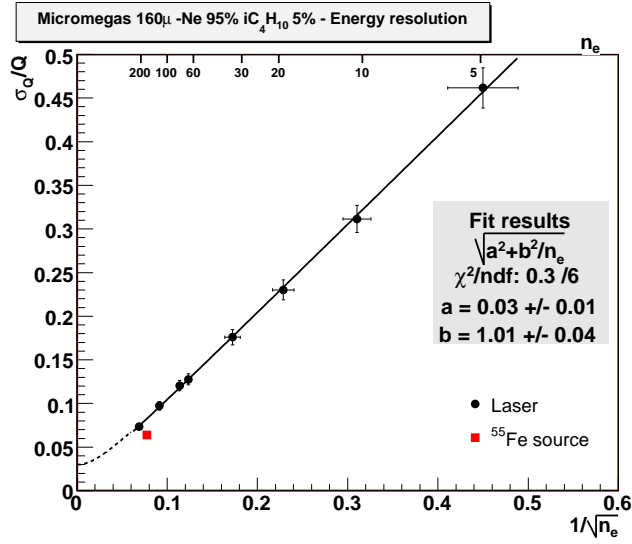}
	\caption{Energy resolution as a function of the number of primary electrons produced by focusing the laser on the drift electrode and varying the light intensity. The data points are fitted following a quadratic function (plain line). The dotted line is an extrapolation of the fit. The point corresponding to the  $\mathrm {^{55}}$Fe source is added on the graph. }
	\label{fig:figure5-eresol-vs-sqrt-ne}
\end{figure}
\newpage
%
%
% TABLE 2
%
%
\begin{table}[htbp]
\centering
\begin{tabular}{|c|c|c|c|}
\hline
Number of primary electrons $\mathrm{n_{e}}$ & Mesh voltage (V) & Gain & Charge resolution \\
\hline
5$\pm$1 & 470 & 3.7 10$\mathrm{^4}$ &0.46 \\
 & 480 & 5.0 10$\mathrm{^4}$ & 0.46 \\
 & 490 & 7.0 10$\mathrm{^4}$ & 0.44 \\
\hline
10$\pm$1 & 440 & 1.4 10$\mathrm{^4}$ & 0.32 \\
& 470 & 3.7 10$\mathrm{^4}$ &0.35 \\
 & 480 & 5.0 10$\mathrm{^4}$ & 0.34 \\
 & 490 & 7.0 10$\mathrm{^4}$ & 0.32 \\
\hline
20$\pm$2 & 440 & 1.4 10$\mathrm{^4}$ & 0.23 \\
& 470 & 3.7 10$\mathrm{^4}$ &0.23 \\
 & 480 & 5.0 10$\mathrm{^4}$ & 0.22 \\
\hline
33$\pm$3 & 440 & 1.4 10$\mathrm{^4}$ &0.17 \\
 & 470 & 3.7 10$\mathrm{^4}$ & 0.19 \\
\hline
\end{tabular}
\caption{Charge resolution for several number of primary electrons at different gains}
\label{table:tab2}
\end{table}
An example of charge spectrum obtained with 33$\pm$3 primary electrons is shown in Fig. \ref{fig:figure4-qtot-ne33-vg470}. The formula (\ref{eqn:chargeresollas}) was used to fit the charge resolution versus the number of injected electrons, as shown in figure \ref{fig:figure5-eresol-vs-sqrt-ne}. The actual fit function, which is shown in the figure inset, uses the parameters $a$ and $b$, which stands for $\sqrt{f+F_{las}}$ . A very good agreement is observed. A value of 0.71 $\pm$ 0.10 for $F_{las}$ is deduced from the fit and the value of $f$ determined from the single electron response measured previously. The $a$ value (3 $\pm$ 1~\%) is consistent with the laser fluctuations measured with the control PMT (2~\%), as well as the value given by the manufacturer ($\leq$ 4~\%). Its contribution is negligible on the main part of the studied range.
Once the laser Fano factor has been determined, the dependence of the energy resolution from the mesh voltage was investigated. Table \ref{table:tab2} sums up the results. It shows that the energy resolution is independent of the gain in the voltage range of operation. This result is consistent with the SER measurements presented in section \ref{sec:single}.

%
%
% SUBSECTION 5.2: MEASUREMENT  WITH THE 55Fe source
%
%
\subsection{Measurement with a $^{55}$Fe source}
\label{subsec:fanosource}

\begin{figure}[htbp]
	\centering
		\includegraphics[width=9cm]{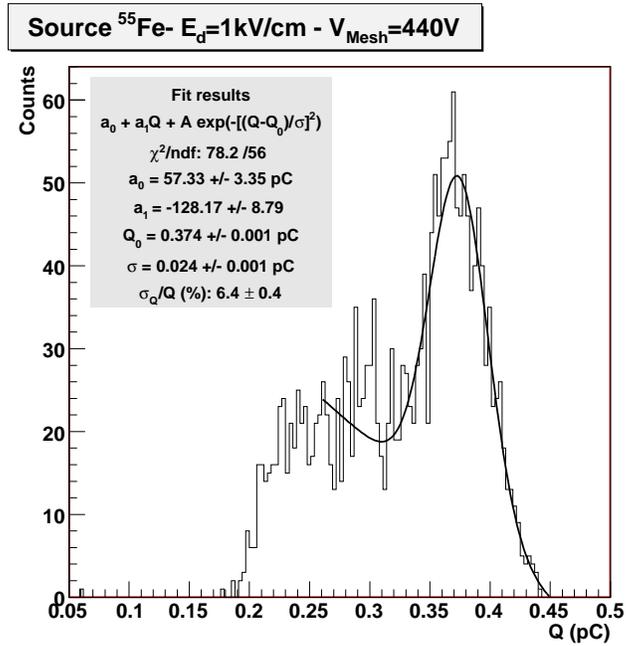}
	\caption{Charge induced on one pad by 5.9~keV X-ray of a  $\mathrm {^{55}}$Fe source. The gas mixture is Ne 95\% iC$_{4}$H$_{10}$ 5\% and the mesh voltage of 440~V. The spectrum is fitted following a function (plain line), which superposes a linear function (background) and a Gaussian (source peak). }
	\label{fig:figure6-xray-spectrum}
\end{figure}

In the case of a particle impinging the gas, the average number of primary electrons is defined by:
$$
\mathrm{E}(n_e)=\frac{\mathcal{E}}{W}
$$
where $W$ is the mean ionisation energy and $\mathcal{E}$ the energy of the incident particle (see section \ref{sec:gaincalib} for the determination of their values). Then, the expression of the energy resolution measured with a $\mathrm{^{55}}$Fe source is obtained by applying (\ref{eqn:fanosource}) to (\ref{eqn:varqeq2}): 
\begin{equation} \label{eqn:chargeresolsource}
\left(\frac{\sigma_Q}{Q} \right)= \frac{\sqrt{\mathrm{var}(Q)}}{\mathrm{E}(Q)}=\sqrt{(f+F)\frac{W}{\mathcal{E}}}
\end{equation}
The charge spectrum measured on the central 4$\times$4-mm$^2$ pad at a mesh voltage of 440~V is shown in Fig. \ref{fig:figure6-xray-spectrum}. The events were selected by anticoincidence with the neighbouring pads. The value of the relative gain variance $f$ for a mesh bias of 440~V is the same than at 470, 480 and 490~V, as shown in section \ref{subsec:fanolaser}. 
From the measured charge resolution, the value of the Fano factor deduced is 0.37. Due to the low counting rate on the central pad, the data had to be acquired over almost 10 hours, and the gas gain could change during the experiment.  If such a variation occurs, the actual value of $f$ is increased and the measured spectrum is wider than expected. Thus the value given for the Fano factor can only be interpreted as a maximum value: $F \leq$ 0.37. This result is a first experimental estimation of this parameter for a binary mixture consisting of Neon and a molecular gas.
%
%
% SECTION: CONCLUSION
%
%
\section{Conclusion}
\label{sec:conclusion}

We succeeded in separately measuring the contribution to the energy resolution of the gain and the ionisation fluctuations of a Micromegas detector. The method relies on a test-bench made of a 337-nm wavelength laser and an optical focusing system. The detector was operated in Ne 95~\% iC$\mathrm{_4}$H$\mathrm{_{10}}$ 5~\%.
The single-electron response (SER) was measured for gains 3.7 10$\mathrm{^4}$, 5.0 10$\mathrm{^4}$ and 7.0 10$\mathrm{^4}$. The experimental data were fitted with a Polya distribution. The obtained Polya parameter ( $\theta=2.2 \pm 0.2$) allowed us to deduce the relative gain variance ($f=0.31 \pm$ 0.02) which did not depend on the gain in the range of operation. The laser system enabled us to study the energy resolution as a function of the number of primary electrons $n_e$. A Fano factor could be deduced in the case of the laser-metal interaction: $F_{las}= 0.71 \pm 0.10$, taking the value of $f$ obtained from the SER. A first experimental determination of the maximum value of the Fano factor for the Ne 95\% iC$\mathrm{_4}$H$\mathrm{_{10}}$ 5\% mixture at atmospheric pressure with 5.9~keV X-rays was obtained:  $F \leq$ 0.37 . 

The experimental method presented here could be widely used for various MPGD's geometries and gas mixtures. It could give experimental measurements of the gain variance and the Fano factor, which are fundamental parameters of these detectors. The detector used in the present work showed good performances in terms of energy resolution at atmospheric pressure, thanks to its low gain fluctuations. Further studies of its properties at different pressures and gas mixtures are planned in the future.

%
%
% SECTION: ACKNOWLEDGEMENTS
%
%
\section*{Acknowledgements}
\label{sec:acknow}

The authors would like to thank A. Maroni, C. Thenault and M. Imre who assembled the detector, T. Nguyen and M. Josselin for their work with the electronics. They would also like to dedicate this paper to the memory of Vincent Lepeltier (1942-2008), who provided a significant contribution to the accomplishment of this work.

%%  APPENDIX A
%%
%%
 \appendix
 \numberwithin{equation}{section}
\section{Charge resolution}
\label{app:appa}
The aim of this appendix is to demonstrate the formula (\ref{eqn:varqeq2}) used in section \ref{sec:eneresol}.

It is assumed that there is no attachment of primary electrons during the drift, that the multiplication in the gas has linear behavior, that all multiplications are independent from one another and have the same probability law. In this case, the response to $n_e$ electrons is the sum of $n_e$ single electron responses. Then, the number $Q$ of electrons detected at the anode is:

\begin{equation} \label{eqn:q}
Q= N_{noise}+ \sum_{1\leq k \leq n_{e}} N_{SER, k}
\end{equation}
where $N_{SER}$ is the random variable of electrons collected at the anode for a single photoelectron (single electron response or SER) and $N_{noise}$ the random variable corresponding to the noise. The sum is calculated over a random number of terms, $n_e$. When $n_e=0$ , $Q$ is evaluated only from the noise. The expectation and variance of $Q$ are calculated from equation (\ref{eqn:q}) assuming that the noise is totally independent from the multiplication process:

$$
\mathrm{E}(Q)=\mathrm{E}(N_{noise})+\mathrm{E}(N_{SER})\mathrm{E}(n_e)
$$
$$
\mathrm{var}(Q)=\mathrm{var}(N_{noise})+\mathrm{var}(N_{SER})\mathrm{E}(n_{e})+\mathrm{E}^2(N_{SER})\mathrm{var}(n_e)
$$ 
The demonstration can be achieved with conditional probabilities, which is less demanding on the properties of the distribution than the method of moment generating functions. The average of the noise is usually 0. Let $f$ designate the relative variance of the single electron response:

$$
f=\frac{\mathrm{var}(N_{SER})}{\mathrm{E}^2(N_{SER})}
$$
and define the following variables:

$$
\sigma_I=\sqrt{\mathrm{var}(N_{noise})}
$$
$$
G=\mathrm{E}(N_{SER})
$$
corresponding to the standard deviation of the noise an to the detector gain respectively. Then:
\begin{equation} \label{eqn:relvarq}
\frac{\mathrm{var}(Q)}{(\mathrm{E}(Q))^2}= \frac{f}{\mathrm{E}(n_e)}+ \frac{\mathrm{var}(n_e)}{\mathrm{E}^2(n_e)}+ \left(\frac{\sigma_I}{G} \right)^2 \frac{1}{\mathrm{E}^2(n_e)}
\end{equation}
In the case of a radioactive source, $\mathrm{var}(n_e)/\mathrm{E}^2(n_e)$ can be expressed with the introduction of the Fano factor \cite{fano47, alkha67}:

$$
\frac{\mathrm{var}(n_e)}{\mathrm{E}^2(n_e)}=\frac{F}{\mathrm{E}(n_e)}
$$
In the case of the laser, $\mathrm{var}(n_e)/\mathrm{E}^2(n_e)$ depends on the laser light fluctuations and on the electron emission statistics, which can be expressed by:

$$
\frac{\mathrm{var}(n_e)}{\mathrm{E}^2(n_e)}=a^2+\frac{F_{las}}{\mathrm{E}(n_e)}
$$
The second term $F_{las}/\mathrm{E}(n_e)$ expresses the contribution of the fluctuations in the electron emission process and has been modelled as a deviation from a standard Poisson law by the introduction of the parameter $F_{las}$, which is analogous to a Fano factor. The $a$ parameter is related to the laser light fluctuations. As discussed in Ref. \cite{anisi77}, the electron emission follows a power law with an exponent $p$ of around 2 . For the small variations of the laser light $\alpha$ (around 2~\% in our setup), a linear expansion of the power law leads to multiply $\alpha$ by $p$ and thus $a=p\alpha \simeq 4~\%$. 
Equation (\ref{eqn:relvarq}) then becomes: 

$$
\left(\frac{\mathrm{var}(Q)}{(\mathrm{E}(Q))^2} \right)=a^2+\frac{F_{las}+f}{\mathrm{E}(n_e)}+\frac{\sigma_I^2}{\mathrm{E}^2(n_e)}
$$

%
%
% REFERENCES
%
%


\begin{thebibliography}{99}
\bibitem{sauli97} 
F. Sauli, Nucl. Instr. and Meth. A 386 (1997) 531.
\bibitem{giomataris96}
Y.Giomataris et al., Nucl. Instr. and Meth. A 376 (1996) 29.
\bibitem{ketz02}
B. Ketzer, Nucl. Instr. and Meth. A 494 (2002) 142.
\bibitem{koba07}
M. Kobayashi et al., Nucl. Instr. and Meth. A 581 (2007) 265.
\bibitem{kuden08}
Y. Kudenko, Nucl. Instr. and Meth. A (2008), doi:10.1016/j.nima.2008.08.029 .
\bibitem{blank08}
B. Blank et al., Nucl. Instr. and Meth. B 266 (2008) 4606.
\bibitem{nakho08}
M. Nakhostin, Nucl. Instr. and Meth. A (2008), doi:10.1016/j.nima.2008.09.025 .
\bibitem{fano47}
U. Fano, Phys. rev. 72 (1947) 26.
\bibitem{alkha67}
G.D. Alkhazov et al., Nucl. Instr. and Meth. 48 (1967) 1.
\bibitem{hurst78}
G.S. Hurst et al., Nucl. Instr. and Meth. 155 (1978) 203.
\bibitem{kase79}
M. Kase et al., Nucl. Instr. and Meth. 163 (1979) 289.
\bibitem{kase84}
M. Kase et al., Nucl. Instr. and Meth. 227 (1984) 311.
\bibitem{delima82}
E.P. de Lima et al., Nucl. Instr. and Meth. 192 (1982) 575.
\bibitem{hashiba84}
A. Hashiba et al., Nucl. Instr. and Meth. 227 (1984) 305.
\bibitem{pansky96}
A. Pansky, A. Breskin and R. Chechik, J. Appl. Phys. 79 (1996) 8892.
\bibitem{pansky97}
A. Pansky, A. Breskin and R. Chechik, J. Appl. Phys. 82 (1997) 871.
\bibitem{byrne62}
J. Byrne, Proc. R. Soc. Edinburg Sect. A 66 (1962) 33.
\bibitem{genz73}
H. Genz, Nucl. Instr. and Meth. 112 (1973) 83.
\bibitem{zerguer07}
T. Zerguerras et al., Nucl. Instr. and Meth. A 581 (2007) 258.
\bibitem{spectra08}
http://www.spectra-physics.com
\bibitem{anisi77}
S.I. Anisimov, V.A. Benderskii and G. Farkas, Sov. Phys. Usp. 20 (6) (1977) 467.
\bibitem{nxpweb}
http://www.nxp.com
\bibitem{santiard94}
J.-C. Santiard et al., Gasplex a low-noise analogue signal processor for read out of gaseous detectors, CERN-ECP/94-17, 1994.
\bibitem{caenweb}
http://www.caen.it
\bibitem{photonisweb}
http://www.photonis.com
\bibitem{leo94}
W.R. Leo, Techniques for Nuclear and Particle Physics Experiments, Second Revised Edition, Springer-Verlag (1994) p.131.
\bibitem{photonis02}
Photonis, Photomultiplier tubes: principles and applications (2002), p.4-12.
\bibitem{bondar07}
A. Bondar et al., Nucl. Instr. and Meth. A 574 (2007) 493.
\bibitem{buzulu00}
A. Buzulutskov et al., Nucl. Instr. and Meth. A 443 (2000) 164.
\bibitem{vavra02}
J. Va'vra and A. Sharma, Nucl. Instr. and Meth. A 478 (2002) 235.
\bibitem{derre00}
J. Derr\'e et al., Nucl. Instr. and Meth. A 449 (2000) 314.
\bibitem{bronic96}
I.K. Bronic and B. Grosswendt, Nucl. Instr. and Meth. B 117 (1996) 5. 
\bibitem{knoll89}
G.F. Knoll, Radiation detection and measurement, Second edition, Wiley and Sons (1989), p.176.
\bibitem{koba06}
M. Kobayashi, Nucl. Instr. and Meth. A 562 (2006) 136.
\bibitem{agraw88}
P.C. Agrawal and B.D. Ramsey, Nucl Instr. and Meth. A 273 (1988) 331.
\end{thebibliography}
\end{document}